\documentclass[sn-mathphys-num]{sn-jnl}

\usepackage[OT4]{fontenc}
\usepackage[utf8]{inputenc}
\usepackage{graphicx}%
\usepackage{multirow}%
\usepackage{amsmath,amssymb,amsfonts}%
\usepackage{amsthm}%
\usepackage{mathrsfs}%
\usepackage[title]{appendix}%
\usepackage{calligra}
\usepackage[dvipsnames]{xcolor}
\usepackage{textcomp}%
\usepackage{manyfoot}%
\usepackage{booktabs}%
\usepackage{algorithm}%
\usepackage{algorithmicx}%
\usepackage{algpseudocode}%
\usepackage{listings}%
\usepackage{natbib}
\usepackage[normalem]{ulem} 
\usepackage{caption}
\usepackage{subcaption}
\usepackage{tikz}
\usepackage{longtable}
\usepackage{dsfont}
\usepackage{datetime}

\usepackage{bm}
\usepackage{csquotes}

\usetikzlibrary{automata, positioning}

\newcommand{\R}{\mathbb{R}}

\newcommand{\diff}{\mathop{}\!\mathrm{d}}

\newcommand*{\secref}[1]{Section~\ref{#1}}

\newcommand*{\subsecref}[1]{Subsection~\ref{#1}}

\begin{document}

\title[Article Title]{Beyond Bayesian Inference: The Correlation Integral Likelihood Framework and Gradient Flow Methods for Deterministic Sampling}



\author[1,2]{\fnm{Piotr} \sur{Gwiazda}}\email{pgwiazda@mimuw.edu.pl}
\author[3]{\fnm{Alexey} \sur{Kazarnikov}}\email{kazarnikov@gmail.com}
\author[3]{\fnm{Anna} \sur{Marciniak-Czochra}}\email{anna.marciniak@iwr.uni-heidelberg.de}
\author[2]{\fnm{Zuzanna} \sur{Szymańska}}\email{z.szymanska@icm.edu.pl}

\affil[1]{\orgdiv{Institute of Mathematics}, \orgname{Polish Academy of Sciences}, \orgaddress{\street{ul.~Śniadeckich~8}, \postcode{00-956}, \city{Warsaw}, \country{Poland}}}
\affil[2]{\orgdiv{Interdisciplinary Centre for Mathematical and Computational Modelling}, \orgname{University of Warsaw}, \orgaddress{\street{ul. Tyniecka 15/17}, \\ \postcode{02-630}, \city{Warsaw}, \country{Poland}}}
\affil[3]{\orgdiv{Institute for Mathematics}, \orgname{Heidelberg University}, \orgaddress{\street{Im Neuenheimer Feld~205}, \postcode{D-69120}, \city{Heidelberg}, \country{Germany}}}

\abstract{
Calibrating mathematical models of biological processes is essential for achieving predictive accuracy and gaining mechanistic insight. However, this task remains challenging due to limited and noisy data, significant biological variability, and the computational complexity of the models themselves. In this method's article, we explore a range of approaches for parameter inference in partial differential equation (PDE) models of biological systems. We introduce a unified mathematical framework, the Correlation Integral Likelihood (CIL) method, for parameter estimation in systems exhibiting heterogeneous or chaotic dynamics, encompassing both pattern formation models and individual-based models. Departing from classical Bayesian inverse problem methodologies, we motivate the development of the CIL method, demonstrate its versatility, and highlight illustrative applications within mathematical biology. Furthermore, we compare stochastic sampling strategies, such as Markov Chain Monte Carlo (MCMC), with deterministic gradient flow approaches, highlighting how these methods can be integrated within the proposed framework to enhance inference performance. Our work provides a practical and theoretically grounded toolbox for researchers seeking to calibrate complex biological models using incomplete, noisy, or heterogeneous data, thereby advancing both the predictive capability and mechanistic understanding of such systems.}

\keywords{parameter estimation, Correlation Integral Likelihood, Bayesian inference, gradient flows, deterministic sampling}


\maketitle


\begin{center}
  {\calligra
  \large\textbf{Dedication}
  \par\vspace{0.3cm}}
\end{center}
\textit{It is our pleasure to dedicate this paper to our friend, Professor Mark A. J. Chaplain, who has been, for many of us, not only an inspiring example of scientific excellence but also a generous mentor and a trusted friend. His unwavering support, thoughtful guidance, and collegial spirit have shaped our work, strengthened our community, and left a lasting impact on both our professional paths and personal lives.}

\textit{Inspired by his remarkable attitude, we pledge that, so long as a hundred of us remain scientifically active, never will we, on any condition, yield the mathematical biology to the lordship of other fields. It is in truth not for glory, nor riches, nor honours that we labour, but for true knowledge alone, which no honest scientist gives up but with life itself.}\\


\section{Introduction}\label{Sec1:Introduction}


In recent years, biology and medicine have been undergoing a fundamental transformation from qualitative, data-poor sciences to quantitative, data-driven disciplines. Advanced technologies have enabled the study of biological systems with unprecedented precision, allowing researchers to collect vast amounts of data. In some areas, such as genome-wide transcription studies, datasets capture nearly all cellular responses to specific stimuli. Nonetheless, data availability remains limited in other instances, offering only dynamic information on protein levels or static snapshots of the spatial structure of solid tumours. Broadly speaking, two modelling paradigms are emerging, built on contrasting methodologies. Statistical and machine learning methods dominate the first paradigm, typically requiring no prior knowledge of the underlying system, as data-driven hypotheses are inferred directly from the data. In contrast, the second paradigm relies on mechanistic models, which require substantial prior knowledge of the system to manually construct representations that capture its essential dynamics. While this approach offers a significant advantage by explaining underlying mechanisms and providing a foundation for controlling the system, it also poses major challenges, as it involves many qualitatively distinct research methodologies.
 
Mathematical models should, of course, be analytically well-posed, demonstrating properties such as existence, uniqueness, and regularity of solutions.
While these properties are important, they are not sufficient for real-world applications. Appropriate illustration of the results, achieved through numerical simulations, is also necessary. These simulations not only facilitate interdisciplinary communication but also enable meaningful comparisons between numerical results and experimental observations. A critical assumption in simulations is that all parameters are known. But is this truly the case? Clearly, it is not, as for many parameters, direct experimental measurement is not possible due to physical unfeasibility or the lack of established experimental protocols. Moreover, even when data collection is feasible, it is often prone to error, yielding probability distributions rather than precise values. Consequently, some parameters must be inferred from indirect measurements, leading to the inverse problem of parameter identification. 

Once again, it becomes evident that practical applications give rise to theoretical challenges. Investigating the regularity of solutions with respect to the model parameters provides information on how variations in parameters lead to changes in the solution of the PDE. If small changes in parameters have a significant impact on solutions, then the inference process may become highly sensitive to noise in the observational data. Conversely, if the model exhibits robustness to variations in parameters, the inference process will be more stable and reliable~\cite{Bianchini2002,debiec2025}. As a reward for that effort, we obtain parameter estimation and model validation, which indicate whether the model accurately represents the real process. A thorough understanding of the mathematical structure of the model and its properties, such as the regularity of solutions with respect to model parameters, is also necessary for another key step in developing truly predictive models, which is the construction of reliable numerical algorithms~\cite{Gwiazda2023}. More precisely, the numerical scheme should satisfy the same stability estimates as the analytical solution of the equation. Finally, effective algorithms in Bayesian statistics, as well as in machine learning, require improvements in sampling methods, which involve the selection of representative data points from a larger data set to estimate the characteristics of the entire population. In the context of parameter estimation, the problem of sampling a probability distribution is of particular interest. 
Employing methods grounded in gradient flows in the space of probability measures and integrating them into a biological inference framework emerges as a promising direction~\cite{Ambrosio_Gigli_Savare_2008}.

The above sketch roughly illustrates the number of challenges scientists face when performing a formal model calibration. It is therefore understandable why most models of biological dynamics, especially in spatially or physiologically structured contexts, have traditionally relied on theoretical assumptions rather than empirical calibration. However, perhaps due to the increasing availability of computational power needed to perform calibration, this situation is beginning to change. Recent studies demonstrate that data-driven model calibration can reveal novel insights into biological mechanisms and is an essential prerequisite for these models to function as genuine tools for model-based data analysis.

Exemplary studies span both discrete and continuous frameworks. In 2024, approximate Bayesian computation was used to calibrate single- and biphasic agent-based tumour growth models, enabling the identification of key parameters such as proliferation rates and spatial constraints on cell division that distinguish tumour growth phases~\cite{Wang2024}. In a paper published the same year, the identifiability of several Fisher–KPP variants for cell invasion was assessed using profile likelihoods, revealing that more complex models, while biologically richer, are often less identifiable and require more data for reliable parameter estimation~\cite{Liu2024}. A year earlier, Bayesian calibration was applied to parameterise a multiphase porous-media tumour model, combining global sensitivity analysis with Bayesian methods to indirectly characterise mechanical properties of neuroblastoma spheroids that are otherwise difficult to measure~\cite{Hervas-Raluy2023}. In 2023, approximate Bayesian computation and gradient matching were compared in a PDE model of cancer spread, demonstrating that approximate Bayesian computation (ABC) is more robust to measurement noise and therefore better suited for parameter recovery in noisy biological data~\cite{Xiao2021}. In the same year, calibrated diffusion–proliferation coefficients were investigated in a murine glioma model, helping to explain the large inter-subject variability in tumour growth by testing hypotheses about diffusion and proliferation dynamics~\cite{Lima2021}. Earlier, in 2017, calibration and validation of Gompertzian spheroid growth were performed, demonstrating a practical workflow for uncertainty quantification and improving predictive accuracy under noisy data conditions~\cite{Collis2017}. Notably, even elementary mechanistic models, when systematically calibrated to data, can yield novel biological insights. For instance, in adult neurogenesis, mathematical models calibrated to experimental data on the dynamics of distinct cell subpopulations have challenged prevailing hypotheses by showing that the age-related decline in neurogenesis is primarily due to reduced activation rates of quiescent neural stem cells rather than changes in their self-renewal probabilities, thereby revising the understanding of how stem cell dynamics change with ageing~\cite{Kalamakis2019,Ziebell2018}. Together, these studies underscore the transformative potential of parameter identification, not only for improving predictive accuracy but also for uncovering hidden mechanisms and reshaping biological understanding.

This paper discusses the potential of new mathematical methods to address a key challenge in the practical application of mathematical models: the lack of proper calibration. It outlines the core ideas behind various sampling techniques, ranging from well-established methods such as the Metropolis and Metropolis-Hastings algorithms to more advanced approaches designed to mitigate the limitations of existing solutions. Building on this foundation, we present the recently developed Correlation Integral Likelihood (CIL) method as a unifying framework tailored for systems characterised by spatial heterogeneity, stochasticity, or chaotic behaviour. We also discuss how deterministic gradient flow methods, when integrated within the proposed framework, may enhance the efficiency and accuracy of inference. In summary, we present a practical and theoretically robust toolkit for calibrating complex biological models using incomplete, noisy, or heterogeneous data by combining stochastic and deterministic inference strategies within the CIL framework, thereby providing a flexible approach to parameter estimation that enhances both predictive power and mechanistic understanding. To guide the reader, the remainder of this article is organised as follows. In \secref{sec:inv_problems}, we review the foundations of parameter identification formulated as inverse problems, highlighting the challenges that arise in complex biological systems. \secref{sec:models_with_heterogenity} introduces our CIL framework, beginning with its application to chaotic dynamical systems (\subsecref{subsec:CIL_approach}) and extending to spatially extended systems, including pattern formation and agent-based models (\subsecref{subsec:application_CIL}). \secref{sec:sampling} explores sampling strategies for parameter inference, comparing stochastic Markov Chain Monte Carlo (MCMC) methods (\subsecref{subsec:MCMC_methods}) with deterministic gradient flow approaches (\subsecref{subsec:gradient_flow}). Finally, \secref{sec:discussion} discusses the implications of our findings and outlines future directions for advancing inference methods in mathematical biology.


\section{Parameter Identification via Inverse Problems \label{sec:inv_problems}}


Inverse problems can be tackled either with deterministic methods, which minimise a predefined cost function, or within a broader stochastic framework, such as Bayesian methods, in which model parameters are treated as random variables described by probability distributions. The key concept is the posterior distribution, which represents the conditional distribution of the parameters given the observations, allowing inferences to be made about individual parameter values. By Bayes' theorem, it is given by
\begin{equation}\label{posterior}
\pi(\theta|D)=\frac{\pi(\theta)\ell(D|\theta)}{\int_{\Theta}\pi(\theta)\ell(D|\theta)d\theta}\,,    
\end{equation}
where $D$ denotes the collected data, $\theta$ represents a given vector of parameters, and $\Theta$ is the space of all parameters. The posterior probability density, $\pi(\theta \mid D)$, represents the probability density of $\theta$ given the data $D$. The prior probability density, denoted by $\pi(\theta)$, is specified based on some prior knowledge, while the likelihood function, $\ell(D \mid \theta)$, quantifies the probability of observing the data $D$ given the parameter~$\theta$.

Suppose we are given a theoretical model that describes the dynamics of a process using a system of differential equations. The objective is to estimate an unknown parameter vector $\theta \in \mathbb{R}^d$ for this model from observed data $D \in \mathbb{R}^k$, within a probabilistic framework grounded in Bayes’ theorem. More precisely, under certain assumptions, the relationship between parameter vector $\theta$ and data $D$ is given by:  
\begin{equation}\label{BI}  
    D = \mathcal{M}(\theta) + \eta,  
\end{equation}  
where $\mathcal{M}: \mathbb{R}^d \to \mathbb{R}^k$ is the forward operator that maps the parameters $\theta$ to the observation space, while $\eta$ represents the observational noise. We assume that the prior probability density $\pi(\theta)$ follows a Gaussian distribution, that is, $\theta \sim \mathcal{N}(m, \Sigma)$. Under this assumption, the prior density is given by
\begin{equation}
    \pi(\theta) = e^{-\frac{1}{2} |\theta-m|^2_{\Sigma}}.
\end{equation}
The term $\frac{1}{2} |\theta - m|^2_{\Sigma}$ represents prior knowledge about $\theta$, penalizing deviations from the prior mean $m$, scaled by the prior covariance $\Sigma$. Now, suppose that the noise $\eta \sim \mathcal{N}(0, \Gamma)$ represents measurement uncertainty. Then the likelihood of observing $D$ given parameters $\theta$ is 
\begin{equation}
    \ell(D|\theta) = e^{-\frac{1}{2} |D - \mathcal{M}(\theta)|^{2}_{\Gamma}}.
\end{equation}
Similarly, it penalizes deviations between the model predictions $\mathcal{M}(\theta)$ and the actual observations $D$, scaled by the noise covariance $\Gamma$, where  
\begin{equation}
   |D - \mathcal{M}(\theta)|^{2}_{\Gamma} = (D - \mathcal{M}(\theta))^T \Gamma^{-1}(D - \mathcal{M}(\theta)). 
\end{equation}
The above leads to the posterior density $\pi(\theta|D)$, which takes the form
\begin{equation}\label{like2}  
    \pi(\theta|D) \propto e^{-V(\theta)}, \quad \text{where} \quad V(\theta) = \frac{1}{2} |D - \mathcal{M}(\theta)|^{2}_{\Gamma} + \frac{1}{2} |\theta - m|^2_{\Sigma}.
\end{equation}  

To demonstrate the procedure of parameter identification using the inverse problem approach, we apply it step by step to a specific example: the mathematical model of cancer colony dynamics studied in~\cite{Szymanska2021,Gwiazda2023}, which is based on \textit{in vitro} experiments~\cite{Folkman1973}. In these experiments, the growth of multicellular spheroid cultures was monitored by measuring their diameters at discrete time points $t_i$. The proposed mathematical model, formulated as an integro-differential equation, captures the dynamics of cancer cell colony expansion. The objective is to calibrate the model with experimental data and develop a quantitative tool capable of reliably predicting colony growth. Consider 
\begin{equation}\label{non-local_proliferation}
\partial _t n(x,t) \ =\  \alpha\, k* n(x,t) \,\Bigl( 1 \ -\  n(x,t) \Bigr),
\end{equation}
where $n(x, t)$ denotes the density of cancer cells at time $t$ and location $x$. The parameter $\alpha$ represents the proliferation rate, and $k = k(x)$ is a time-independent, radially symmetric kernel function with compact support, ensuring that 
\begin{displaymath}
k \ast n(x,t) \ =\ \int_{\R^3} k(x-y) n(y,t)\;dy. \;\; 
\end{displaymath}
It is assumed that $k(x)$ has a profile $K$, meaning that $k(x) = K(|x|)$ for $x \in \mathbb{R}^3$. The normalised characteristic function of a ball is adopted as the kernel profile
\begin{equation}\label{kernel}  
K\big(|x|\big) = \frac{3}{4\pi} \sigma_k^{-3} \, \mathds{1}_{\left[0,\sigma_k\right]}\big(|x|\big),  
\end{equation}  
where $\sigma_k$ represents the kernel size (radius). To highlight the core idea, the initial condition is simplified relative to the original article. Specifically, it is assumed to be the characteristic function of a ball, where 
$\sigma_s$ denotes the colony radius at the start of the measurements. Therefore, we take
\begin{equation}\label{initial_condition}
p(r,0) = 4 \pi r^2 \mathds{1}_{\left[0,\sigma_s\right]}(r).
\end{equation}
The inverse problem approach requires linking the model equations to the observed data-specifically, in our case, diameter (or equivalently, radius) measurements, and typically involves extensive simulations. We exploit the characteristics of the experimental data and the assumed radial symmetry of the kernel to reformulate the model in spherical coordinates~\cite{Szymanska2021,Gwiazda2023}. This transformation reduces the problem’s dimensionality, yielding a theoretical framework in which $p_{\theta}(r, t)$ describes the radial density dynamics of the colony, with $r(t)$ representing the colony radius, $t$ denoting time, and $\theta$ being a vector of model parameters, that is, $\theta = [\alpha, \sigma_s, \sigma_k]$. An additional benefit of this change of variables is improved computational efficiency. While the concept of a radius is intuitive, it must be precisely defined for parameter estimation, especially in the context of Eq.~\eqref{non-local_proliferation}, which does not preserve compact support. Measurements of spheroid diameters, and thus their radii, inevitably contain errors. Assuming a homogeneous mass distribution and that the colony radius defines a sphere enclosing 95\% of the total mass, we derive the following expression
\begin{equation}\label{promien}
  \mathcal{M}(\theta) = r_{\theta}(t) = \inf \left\{ q: \int_{0}^q p_{\theta}(r,t)\,dr > 0.95 \cdot \int_{0}^{\infty} p_{\theta}(r,t)\, dr\right\}, 
\end{equation}
where $\int_{0}^{\infty} p_{\theta}(r,t)\,dr$ corresponds to the colony mass at time $t$. 

Another modelling choice is the error model. We assume deviations in measured spheroid diameters scale proportionally to their true values, which is a multiplicative error model, given by
\begin{equation}\label{m_obs}
    r_{o}^i = r_{\theta}(t_i) \cdot Z_i, \;\;\; \textrm{with} \;\;\; \log (Z_i) \sim N(0,\sigma_{o}^2),
\end{equation}
or alternatively
\begin{equation}\label{m_obs_log}
    \log (r_{o}^i) = \log (r_{\theta}(t_i)) + \tilde{Z}_i, \;\;\; \textrm{with} \;\;\; \tilde{Z}_i \sim N(0,\sigma_{o}^2),
\end{equation}
where $r_o^i$ is the measured colony radius at time $t_i$, $r_{\theta}(t_i)$ is the actual colony radius at that time given by Eq.~\eqref{promien}, and $\sigma_o$ represents the measurement error, assumed to be homogeneous over time. 

As the next step in the procedure, we extend the parameter vector associated with the theoretical model Eq.~\eqref{non-local_proliferation} by including the parameter $\sigma_o$, yielding $\theta := \theta \oplus \sigma_o$. The resulting vector of unknown parameters is $\theta = \big[\alpha, \sigma_{k}^2, \sigma_{s}^2, \sigma_{o}^2\big]$, where each component corresponds to the proliferation rate, kernel size, initial colony radius, and measurement error, respectively. A direct consequence of assumption Eq.~\eqref{m_obs} is that the likelihood function $\ell(D \mid \theta)$ is given by
\begin{equation}\label{likelihood}
    \ell(D \mid \theta) = \prod_i  \frac{1}{\sqrt{2 \pi}\sigma_{o}} \exp{\left( \frac{-\Big(\log(r_{o}^i)-\log(r_{\theta}(t_i))\Big)^2}{2 \sigma_{o}^2}\right)},
\end{equation}
where $r(t_i)$ is the colony radius at time $t_i$ for vector of parameters $\theta$.  For convenience, and to avoid imposing positivity constraints, we work with the logarithms of the parameters rather than the parameters themselves. By evaluating the likelihood Eq.~\eqref{likelihood} and assuming a prior distribution $\pi(\theta)$, we obtain the posterior probability density Eq.~\eqref{posterior}. However, computing the posterior can be computationally expensive, and considerable effort has been devoted to optimising the cost of its evaluation.


\section{Parameter inference in models with emergent heterogeneity}\label{sec:models_with_heterogenity}


A central challenge in data-driven model identification arises when limited data must be used to infer parameters in systems that exhibit solution heterogeneity. This is particularly evident in models such as pattern formation models and individual-based models, where complex spatial structures emerge from underlying dynamics. In many cases, available data are restricted to single spatial snapshots, providing limited information about the temporal processes that drive structure formation. Moreover, these heterogeneous outcomes are often highly sensitive to initial conditions, which are typically unknown or poorly constrained.

For example, in two-dimensional Turing models, even small perturbations of the initial conditions can lead to markedly different patterns. Mixed-mode patterns, such as combinations of spots and stripes, may emerge from the same parameter set, reflecting the system’s sensitivity to initial states. The limited availability of data makes it especially challenging to identify and interpret such complex behaviours. Another illustrative case in mathematical biology is stochastic modelling, particularly in individual-based models, where outcomes inherently vary between simulation runs. Here, random processes govern the behaviour of individual agents or entities, so even with identical initial conditions and parameters, repeated simulations can yield different results.

To address this challenge, we present a recent approach known as the Correlation Integral Likelihood method. The name reflects its foundation in the concept of correlation dimension, originally developed for characterising chaotic attractors~\cite{Grassberger1983, Grassberger2007}. Rooted in nonlinear dynamics and chaos theory, the correlation dimension provides a means of estimating the fractal dimension of a point cloud generated by simulations of chaotic systems. Building on and generalising this concept~\cite{Haario2015}, the CIL method constructs a stochastic cost function for parameter estimation in chaotic dynamical systems. This framework is broadly applicable to models with intrinsically random or heterogeneous outcomes. Initially developed for chaotic dynamical systems~\cite{Haario2015, Springer2019, Springer2021, Shah2023, Fumagalli2025}, the CIL method has since been successfully extended to a variety of other model classes. These include cellular automata models~\cite{Kazarnikov2023a} as well as partial differential equation (PDE) models of pattern formation~\cite{Kazarnikov2020, Kazarnikov2023, Pan2025} and of rumour propagation, in both continuous and network-based spatial domains~\cite{Zhu2022, Li2024, Tao2025}. More recently, the method has been adapted to address scenarios with limited data through the introduction of a synthetic variant, known as the Synthetic Correlation Integral Likelihood (SCIL)~\cite{Kazarnikov2023}. Additionally, the method has been extended to handle mixed mode patterns by constructing an ensemble of synthetic likelihoods, thereby capturing the variability of coexisting pattern types~\cite{Kazarnikov2025}.


\subsection{The CIL approach for chaotic dynamical systems}\label{subsec:CIL_approach}


Chaotic dynamical systems are characterised by extreme sensitivity to initial conditions, where even small changes in the initial state or solver parameters can lead to drastically different trajectories, a phenomenon commonly referred to as the butterfly effect. Despite these divergent paths, the trajectories remain confined to the same underlying attractor, typically exhibiting fractal geometry. As a result, there is no single model trajectory corresponding to fixed parameters; instead, there exists an ensemble of trajectories that share the same global geometric properties. This can be understood as the system's natural or intrinsic variability.

While in the parameter estimation problem, the goal is to fit a model to observed data sampled at discrete time points, the inherent sensitivity of chaotic systems makes conventional methods, such as least squares fitting, ineffective, as they rely on pointwise comparisons between data and model predictions. Due to the exponential divergence of trajectories over time, even simulations with identical parameters will quickly diverge from one another because of the system's intrinsic variability. As a result, an alternative metric for model comparison is required. The CIL method addresses these challenges by robustly comparing the global structure of chaotic attractors. It introduces a novel metric based on the concept of generalised correlation integrals~\cite{Grassberger1983}.

\begin{figure}
\centering
\includegraphics[width=0.99\textwidth]{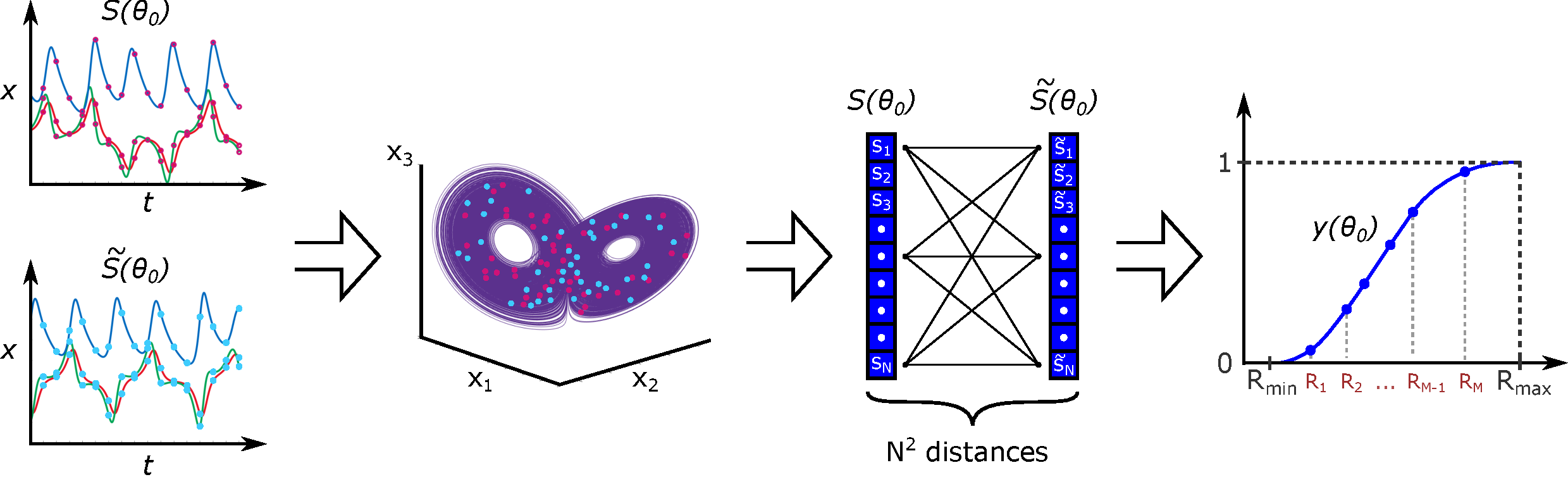}
\caption{
Chaotic dynamical systems display behaviour in which small changes in the initial state, with fixed parameters, result in qualitatively different solutions. The core idea of the CIL method is to map a pair of chaotic trajectories (simulated or observed experimentally) into an eCDF vector of distances between their points. This approach enables the characterisation of the geometric structure of the attractor that both trajectories occupy.}
\label{cil:fig:1}
\end{figure}

Consider a chaotic dynamical system given by
\begin{equation}
\dot{x} = F(x,\theta_0),
\label{cil:chaotic_system}
\end{equation}
where \(x \in \mathbb{R}^n\) is the state vector and \(\theta_0 \in \mathbb{R}^p\) is the vector of parameters. Suppose measurements yield two trajectories of system Eq.~\eqref{cil:chaotic_system}
\[
S(\theta_0) = \{s_1, s_2, \dots, s_N\} \subset \mathbb{R}^n
\]
and
\[
\tilde{S}(\theta_0) = \{\tilde{s}_1, \tilde{s}_2, \dots, \tilde{s}_N\} \subset \mathbb{R}^n,
\]
sampled at time points \(t_1 < t_2 < \cdots < t_N\). Although the trajectories may be generated under slightly perturbed initial conditions or solver settings, they lie on the same attractor as the underlying parameters ${\theta_0}$ are identical. Let \(y(\theta_0) \in \mathbb{R}^M\) be the empirical cumulative distribution function (eCDF) vector, where each component is defined as
\begin{equation}
y_k({\theta_0}) = \frac{1}{N^2} \sum_{i,j=1}^{N} \#\left(\| s_i - \tilde{s}_j \| < R_k\right), \quad k = 1, \dots, M.
\label{cil:eq:eCDF}
\end{equation}
Here, \(\#(\cdot)\) is the indicator function, returning 1 when the distance between the points $s_i$ and $\tilde{s}_j$ is less than 
$R_k$ and 0 otherwise. This vector captures the distribution of distances between points of two trajectories at multiple scales (see Figure~\ref{cil:fig:1}). 

Expression~\eqref{cil:eq:eCDF} is deeply rooted in chaos theory. When \( S(\theta_0) = \tilde{S}(\theta_0) \), the components \( y_k \) estimate the average probability that two states along the trajectory are separated by a distance less than \( R_k \). In this case, the values \( y_k \) are known as \textit{correlation sums}, and their limiting values---obtained as the number of points increases---are referred to as \textit{correlation integrals}. The resulting eCDF vector \( y(\theta_0) \), comprising correlation sums across multiple radii \( R_k \), is therefore termed a \textit{correlation integral vector}.


For small \( R_k \), the correlation sum behaves as \( y_k \sim (R_k)^\nu \), where \( \nu \) denotes the \textit{correlation dimension}, one of the possible ways to measure the dimension of a fractal set~\cite{Grassberger1983}. When estimating this quantity from a finite set of data points, the correlation integral vector is used to determine the optimal value of \( R_k \), ensuring a numerically stable estimate. However, the CIL method takes into account the entire distribution of distances between \textit{different} chaotic trajectories, \( S(\theta_0) \) and \( \tilde{S}(\theta_0) \). For this reason, the eCDF vector \( y(\theta_0) \) is referred to as a \textit{generalised correlation integral vector}.


The variability of the vector \(y(\theta_0)\) reflects the intrinsic chaotic behaviour for fixed parameters $\theta_0$. By repeatedly simulating the system with perturbed initial values, one can empirically estimate the statistics of \(y(\theta_0)\). Numerical experiments show that it can be accurately approximated with a Gaussian distribution. Indeed, if we treat the scalar distances as independent and identically distributed values (i.i.d.), the eCDF vector
\(y(\theta_0)\) converges to a $M$-dimensional Gaussian distribution as $N\rightarrow+\infty$ due to the Central Limit Theorem (CLT), please see \cite{Kazarnikov2025} for details. Naturally, in the case of a chaotic trajectory, measurements are not fully i.i.d. but Gaussianity still holds under suitable conditions on the dependence of the data~\cite{Borovkova2001,Neumeyer2004}. In what follows, we will denote this limiting Gaussian distribution by \( N(\mu(\theta_0), \Sigma(\theta_0)) \), where \(\mu(\theta_0) \in \mathbb{R}^M \)  and \(\Sigma(\theta_0) \in \mathbb{R}^{M \times M} \) denote the mean vector and covariance matrix, respectively.


\begin{figure}
    \centering  \includegraphics[width=0.99\textwidth]{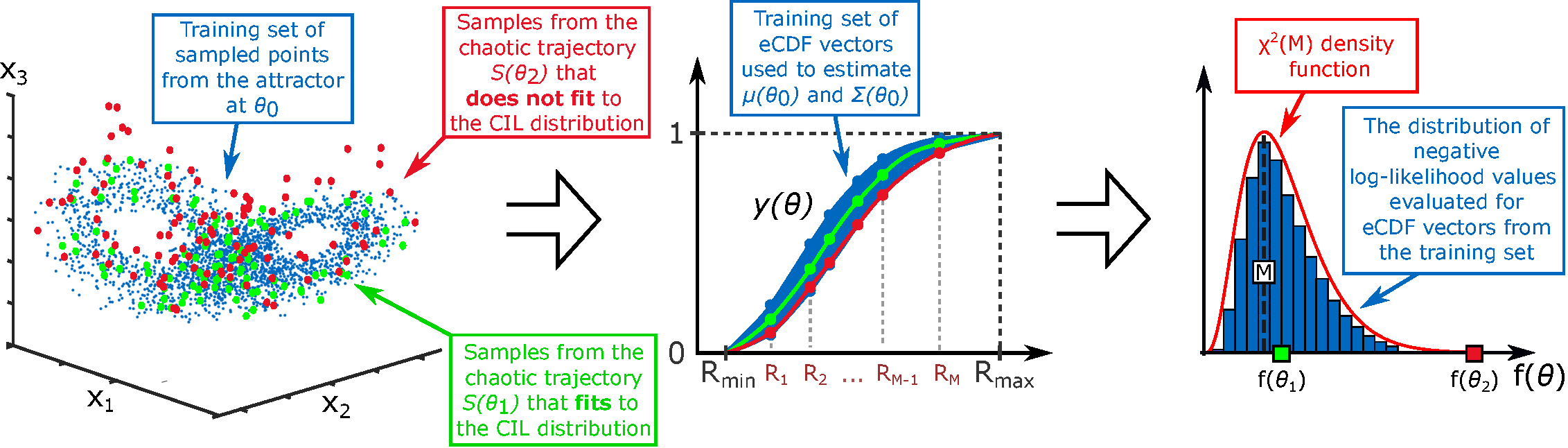}
    \caption{Statistical parameter identification in chaotic systems using the CIL method in the large data case. An experimentally observed trajectory, $S(\theta_0)$, containing a sufficiently large number of data points (shown in blue), is used as a training set to estimate the statistics of a generalized correlation integral vector, $y(\theta_0) \sim N(\mu_0,\Sigma_0)$. Here, $y(\theta_0)$ serves as a characteristic of chaotic behavior, quantifying differences in system dynamics caused by intrinsic chaotic variability. This allows one to compare experimental data with model-generated chaotic trajectories  and measure the level of discrepancy for different values of model parameters. Parameter values that generate trajectories fitting the statistical distribution $N(\mu_0,\Sigma_0)$ are considered equivalent (shown in green), while those producing trajectories that deviate from this distribution are regarded as distinct (shown in red). The statistical distance between a trajectory generated by an arbitrary candidate parameter $\theta$, and the distribution $N(\mu_0,\Sigma_0)$ can be quantified using the standard Gaussian likelihood function. Meanwhile, the range of parameter values that fit to the distribution can be estimated using MCMC sampling methods.}
    \label{cil:fig:2}
\end{figure}

Suppose that we have estimated \( \mu(\theta_0) \) and \( \Sigma(\theta_0) \) for some parameter vector \( \theta_0 \) either from simulated or experimental data. We can evaluate the similarity between the model output for a candidate parameter vector \( \theta \) and the distribution \( N(\mu(\theta_0), \Sigma(\theta_0)) \) using the non-normalised negative log-likelihood function
\begin{equation}
f(\theta) = \bigl(y(\theta) - \mu(\theta_0)\bigr)^\top\,\Sigma(\theta_0)^{-1}\,\bigl(y(\theta) - \mu(\theta_0)\bigr),
\label{cil:cf}
\end{equation}
where the trajectory data for computing a eCDF vector \(y(\theta)\) is created by model simulations for the candidate parameter vector \(\theta\). Equivalently, one can call \( f(\theta) \) the Mahalanobis distance between the vector \( y(\theta) \) and \( N(\mu(\theta_0), \Sigma(\theta_0)) \). Function \( f(\theta) \) is stochastic and from the properties of the Gaussian distribution, it can be deduced that \(f(\theta_0)\) follows the chi-squared distribution with $M$ degrees of freedom: $f(\theta_0) \sim \chi^2(M)$. This idea is illustrated in Figure~\ref{cil:fig:2}.

The parameter identification problem for chaotic systems can be formulated as follows: given an observed trajectory \( S(\theta_0) \), consisting of $N_\text{data}$ measurements, where ${\theta_0}$ is unknown, recover the value of ${\theta_0}$. The CIL method can be applied to this problem in two ways, depending on the amount of available data:

\begin{enumerate}
    \item \textbf{Large Data Case:}  
    If \(N_{\text{data}}\) is sufficiently large, the data can be partitioned into \(n_{\text{ens}}\) subsets, each containing \(N\) measurements, such that \(N^2\) is large enough to compute a robust estimate of 
    the eCDF vector, and \(n_{\text{ens}}\) is sufficient to obtain reliable estimates of \(\mu(\theta_0)\) and \(\Sigma(\theta_0)\) from available $\frac{n_\text{ens}(n_\text{ens} - 1)}{2}$ subset pairs~\cite{Haario2015}. In this case, one can pre-compute these statistical parameters from the observed data \( S(\theta_0) \), and define a stochastic cost function \(f(\theta)\) using expression~\eqref{cil:cf}. Note, that computing \(f(\theta)\) requires creating a trajectory data by simulating model~\eqref{cil:chaotic_system} for the candidate parameter $\theta$, which is needed to define a realisation of a eCDF vector \(y(\theta)\).
    \item \textbf{Limited Data Case:}  
    If \(N_{\text{data}}\) is too small to allow for the above partitioning, the roles of measurements and simulated data can be interchanged. In this variant, for each candidate parameter \(\theta\), repeated model simulations are performed in order to estimate \(\mu(\theta)\) and \(\Sigma(\theta)\) from the simulated data. The observed trajectory \(S(\theta_0)\) is then used to create one eCDF vector \( y(\theta) \) that is used to evaluate the likelihood \( f(\theta) \) for the synthetically created distribution using the expression~\eqref{cil:cf}. This idea is a special case of a Bayesian synthetic likelihood (BSL) method~\cite{Price2018,Wood2010}. Although it is clearly more computationally expensive, it can be used for the limited data case.
\end{enumerate}

In both variants, the CIL method allows for defining a stochastic cost function \( f(\theta) \) for parameter identification that can be minimised using existing methods of stochastic optimisation. Once the local minimum is reached, one can proceed with uncertainty quantification using the MCMC sampling methods.


\subsection{Application of the CIL approach to spatial dynamics: Pattern formation and agent-based models}\label{subsec:application_CIL}


\begin{figure}
     \centering
     \includegraphics[width=0.99\textwidth]{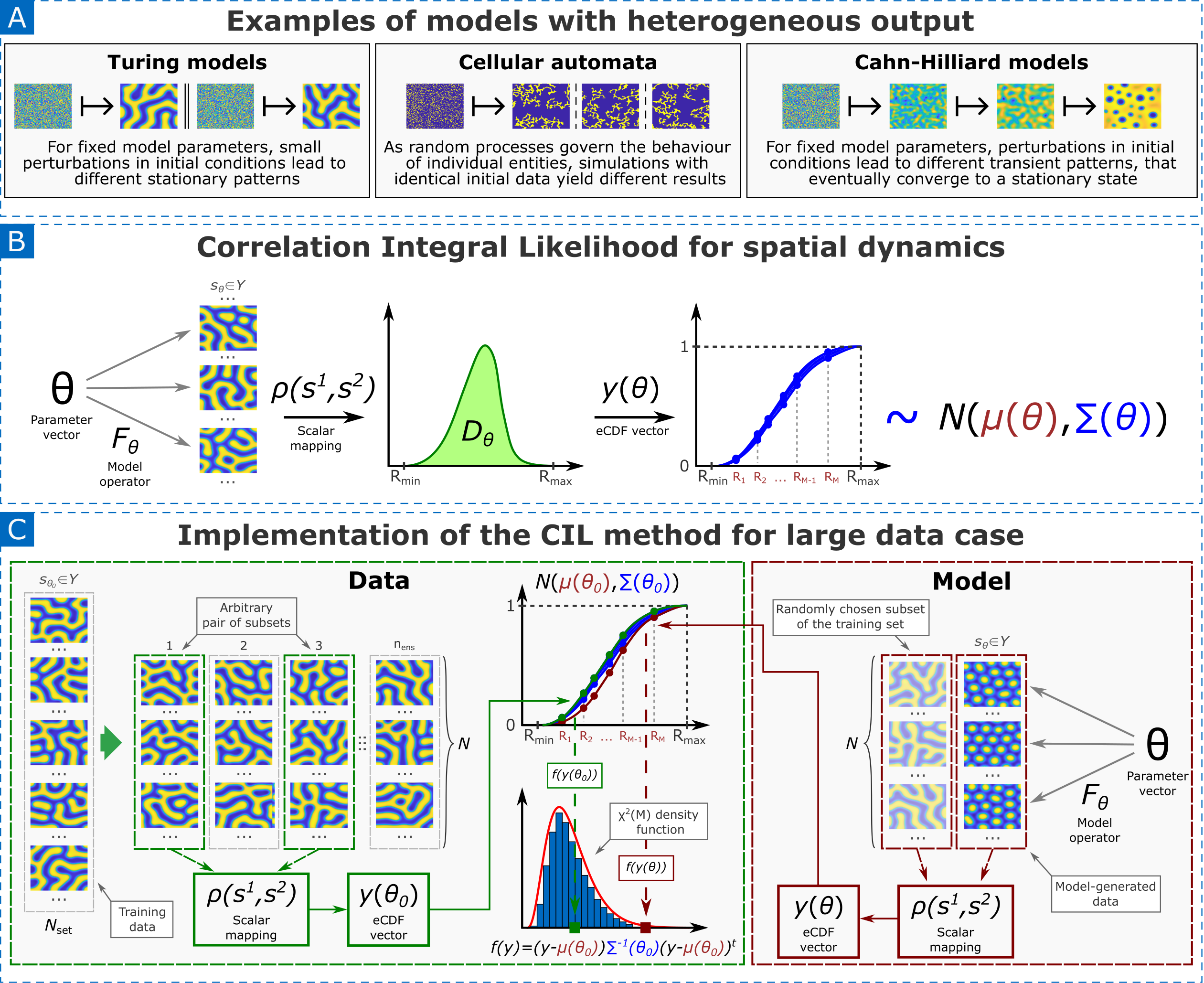}
     \caption{Parameter identification in spatial models with heterogeneous output with Correlation Integral Likelihood method. A: Some examples of spatial models that exhibit heterogeneity in solutions. B: Mapping a family of high-dimensional patterns to the distribution of a scalar random variable (distance between patterns) with the CIL method. C: Parameter identification using the CIL method for the large data case. Here, the multidimensional Gaussian distribution of the eCDF vector $y(\theta_0)$ is derived from experimental data. When a new parameter vector $\theta$ is proposed, one realisation of $y(\theta)$ is created using model-generated (synthetic) data and compared against the reference distribution $N(\mu(\theta_0),\Sigma(\theta_0))$.}
     \label{cil:fig:3}
 \end{figure}

In parameter identification problems for chaotic systems, we are dealing with transient dynamics where multiple observations are collected at different times from a single experiment. In contrast, pattern formation PDE models typically produce a single output per experiment. Despite this difference, both systems produce an ensemble of solutions for fixed parameter values. We will refer to this behaviour as intrinsic \textit{heterogeneity in solutions}.

Several factors contribute to this heterogeneity (Figure~\ref{cil:fig:3}, part A). For instance, in Turing pattern formation models, small perturbations in initial conditions can lead to distinct stationary patterns. Stochastic agent-based models, such as cellular automata, generate varying outcomes even with identical initial settings. In contrast, Cahn-Hilliard models describing the phase separation in binary fluids produce different transient patterns for fixed model parameters due to perturbations in initial conditions, which eventually converge to a stationary state. This inherent variability poses a significant challenge for parameter identification.

In many experimental settings, it is only possible to observe the stationary regime of the pattern formation process, without detailed knowledge of the transient dynamics or the initial state. Furthermore, data are often provided in a normalised form, revealing the structure of the pattern but lacking quantitative concentration levels. In some cases, certain components of the pattern may be inaccessible due to limitations in measuring all relevant variables simultaneously or because the model incorporates hypothetical quantities. Additionally, the observed pattern might not be stationary but could still be transitioning toward a steady state. These constraints make many standard parameter estimation methods unfeasible, as they typically require complete knowledge of all pattern components or assume that the system has reached equilibrium. However, the CIL method remains applicable in such cases, offering a robust solution for parameter identification.


Let us consider an abstract pattern formation model ${s} = {F}_{{\theta}} ({x})$, 
where ${\theta} \in \mathbb{R}^p$ is a vector of parameters, ${x} \in X$ represents the initial data, ${s} \in Y$ is the output pattern, and ${F}_{{\theta}} : X  \rightarrow Y$ for fixed ${\theta} \in \mathbb{R}^p$ is the model operator that maps the initial condition to the final pattern. Since patterns ${s}$ arise from varying initial conditions or stochastic model operators, they can be interpreted as realisations of a random variable. Let us denote the respective distribution of solutions for a fixed parameter set ${\theta} \in \mathbb{R}^p$ by ${S}_{{\theta}}$. Additionally, as numerical experiments typically employ spatially discretised models, we assume that produced patterns are finite-dimensional. 

Let  $\rho : Y \times Y \rightarrow \mathbb{R}_{\geq 0}$ represent a distance between two patterns, based on a suitable norm or metric. To characterise the family of patterns generated by the model ${F}_{{\theta}}$ for a fixed parameter vector ${\theta}$, we introduce the scalar random variable $D_{{\theta}} = \rho({S}^{1}_{{\theta}}, {S}^{2}_{{\theta}})$, where ${S}^{i}_{{\theta}}, i=1,2,$ are two copies of the random variable ${S}_{{\theta}}$, 
induced by two i.i.d. copies of the initial values. This process is illustrated in Figure~\ref{cil:fig:3} (part B). Since patterns are typically bounded, we assume that $D_{{\theta}}$ is also bounded and define 
$R^\text{max}_{{\theta}}  = \text{ess}\,\sup_{{s}_{{\theta}}^1, {s}_{{\theta}}^2} \; \rho({s}^1_{{\theta}}, {s}^2_{{\theta}})$ where ${s}^1_{{\theta}}, {s}^2_{{\theta}}$ are arbitrary realisations of ${S}_{{\theta}}$. 

Consider a set of i.i.d. samples $\{ d^1_{{\theta}}, \ldots, d^n_{{\theta}} \}$ of the random variable $D_{{\theta}}$ obtained by using independent copies of the initial conditions in $n$ runs of the model. We define by ${y}({\theta}) \in \mathbb{R}^M$ 
the eCDF of the distances $d^1_{{\theta}}, \ldots, d^n_{{\theta}}$, which approximates the cumulative distribution function $F_{D_{{\theta}}}$ of  $D_{{\theta}}$ with a finite number of observations. The components $y_k({\theta})$ are given by the formula:
\begin{equation}
y_k({\theta}) = \dfrac{1}{n} \sum \limits_{i = 1}^n \# (d^i_{{\theta}} < R_k), \quad k = 1,\ldots,M,
\label{eq:cil_vector}
\end{equation}
where $0 
< R_1 < \ldots < R_M < R^\text{max}_{{\theta}}$ are fixed threshold values for the bins. As samples $d_\theta^i$, $i=1,\ldots,n$ are obtained from independent model runs ${y}({\theta}) \in \mathbb{R}^N$ asymptotically tends to a multivariate Gaussian distribution as $n\rightarrow +\infty$ due to the CLT~\cite{Kazarnikov2023, Kazarnikov2025}. As before, we will denote the mean and the covariance matrix of the Gaussian distribution by ${\mu}({\theta})$ and ${\Sigma}({\theta})$. 

Although the overall scheme of the CIL method remains almost unchanged compared to the previously discussed case of chaotic dynamical systems, there is one structural distinction. The eCDF vector \eqref{eq:cil_vector} is constructed from distances between spatial patterns, while the expression \eqref{cil:eq:eCDF}, used for chaotic systems, is based on pairwise distances between trajectory points. Despite this apparent difference, the underlying idea remains the same. 

For chaotic dynamical systems, trajectory points are actually samples from the chaotic attractor, which are only weakly dependent provided the time window between them is sufficiently large. In contrast, for pattern formation models, simulated patterns are fully independent because they correspond to distinct realisations of randomised initial conditions. A single simulation of a pattern formation model, however, yields one observation of a pattern, while an integration of a chaotic system produces multiple measurements from the attractor. This distinction leads to substantially higher computational costs in the case of pattern formation models, often necessitating the development of efficient parallel algorithms for the numerical integration of the forward problem.

The problem of parameter identification using pattern data can be formulated as follows: given a finite set of $N_\text{data}$ patterns ${S}_\text{data}$, find parameters ${\theta}_0$ minimising the discrepancy between ${S}_\text{data}$ and pattern family ${S}_{{\theta}_0}$ obtained from repeated simulations of the model ${F}_\theta$. Patterns in the set ${S}_\text{data}$ may come from experimental observations or model simulations. In the latter case,  ${S}_\text{data} =\{{s}^i_{{\theta}_0} = {F}_{{\theta}_0}({x}_i) : i=1,\ldots,N_\text{set}\}$, where ${x}_i$ and ${\theta}_0$ are unknown.
The CIL approach allows us to define a stochastic cost function 
that quantifies the discrepancy between the model ${F}_{{\theta}}$ and the pattern data $S_\text{data}$. The actual implementation of the cost function depends on the amount of data $N_\text{data}$.

If the size  $N_\text{data}$ of the training set is sufficiently large, we can derive enough independent observations of $D_{{\theta}}$ from $S_\text{data}$ to obtain stable estimates for ${\mu}({\theta}_0)$ and ${\Sigma}({\theta}_0)$. We subdivide the training set $S_\text{data}$ into $n_\text{ens}$ subsets, denoted $S^k$,  $k=1,\ldots,n_\text{ens}$, each consisting of $N$ samples of patterns, such that $N_\text{data} = n_\text{ens} \times N$. Next, computing the eCDF vector of the $N \times N$ distances between two sets of $N$ patterns, $S^k$ and $S^l$, for all possible values of $k \ne l$ provides $\left({n_\text{ens} \atop 2}\right)$ realisations of the correlation integral vector ${y}^{k,l}_{0}$, which we use to obtain estimates ${\mu}_0$ and ${\Sigma}_0$ for the parameters of ${y}({\theta}_0)$. In this context, the data is not truly i.i.d, however, in numerical experiments, this approximation closely follows the density of $D_{{\theta}}$ even for relatively small values of $N$~\cite{Kazarnikov2020, Kazarnikov2023, Kazarnikov2025}. This procedure is shown in Figure~\ref{cil:fig:3} (part C).

Once the estimates are obtained, the cost function for parameter estimation can be defined as the non-normalised negative log-likelihood function
\begin{equation}
	f({\theta}) = \big({y}({\theta})-{\mu}_0 \big)^\top {\Sigma}_0^{-1} \big({y}({\theta})-{\mu}_0\big),
	\label{cil:CF}
\end{equation} 
where the correlation integral vector ${y}({\theta})$ at an arbitrary candidate parameter value ${\theta}$ is evaluated by using $N$ simulated patterns obtained from ${F}_{{\theta}}$ and one randomly chosen subset $S^k \subset S_\text{data}$. 

In the large data case, the estimates for ${\mu}_0$ and ${\Sigma}_0$ are constructed off-line from a \enquote{sufficiently large} data set, while the evaluation of $f({\theta})$ during the parameter estimation only requires a \enquote{small} number $N$ of model evaluations. This procedure, combined with bootstrapping, works stably for $N_\text{data} \geq 50$~\cite{Kazarnikov2020}. In the small data case, resampling techniques are insufficient to estimate ${\mu}_0$ and ${\Sigma}_0$ accurately enough from the training set $S_\text{data}$. In that case, we can reverse the roles of the experimental data and the simulations, which leads to the SCIL method mentioned earlier~\cite{Kazarnikov2023, Kazarnikov2025}. Despite being more computationally expensive during parameter estimation than the original CIL approach, the SCIL method can be applied for severely limited pattern data sets, down to the case of a single pattern snapshot.


\section{Sampling}\label{sec:sampling}


In Bayesian inference and uncertainty quantification, sampling methods play a fundamental role in approximating high-dimensional probability distributions. Whether estimating unknown parameters or computing expectations under a posterior distribution, it is often impossible to obtain analytical solutions. As a result, one must resort to numerical techniques that generate representative samples from the target distribution. This chapter reviews both classical and modern approaches to sampling, highlighting their theoretical underpinnings, practical challenges, and recent advances. We begin with stochastic methods based on MCMC, followed by more recent deterministic techniques rooted in gradient flow theory and optimal transport.


\subsection{MCMC methods}\label{subsec:MCMC_methods}


Numerical methods for approximating the joint posterior distribution in Eq.~\eqref{posterior} often rely on MCMC techniques, with the Metropolis and Metropolis-Hastings algorithms being among the most widely used~\cite{mccormick2014,bao2012}. The fundamental idea behind MCMC is to construct a Markov chain whose stationary distribution matches the posterior distribution. After a sufficiently large number of steps, the samples generated by the chain converge to the target distribution given by Eq.~\eqref{posterior}. The process begins by selecting an initial state, $\theta_0$, from a predefined {\it a priori} distribution, $\pi(\theta)$, which encodes prior assumptions about the parameters being estimated. In subsequent steps, the parameter space $\Theta$ is explored by sampling from a probability density function, commonly known as the proposal or jumping distribution. This distribution suggests a candidate $\tilde{\theta}$ for the next sample, $\theta_{j+1}$, based on the previous value, $\theta_j$. When the unknown parameters form a vector, the jumping distribution is often modelled as a multivariate normal distribution, which simplifies the generation of candidate states by leveraging the current state as a reference.  

Among the most widely used implementations of the MCMC framework are the Metropolis algorithm and its more general version, the Metropolis-Hastings algorithm. The primary distinction lies in how they handle the proposal distribution: the Metropolis algorithm assumes symmetry, simplifying the acceptance criterion, while the Metropolis-Hastings algorithm accommodates asymmetry, offering greater flexibility. This adaptability allows for more efficient state space exploration by tailoring the proposal distribution to the specific problem. In practice, the Metropolis-Hastings algorithm is preferred because it incorporates adaptive, problem-specific proposals, improving convergence efficiency. However, when a symmetric proposal distribution suffices, the Metropolis algorithm remains a valid and more straightforward special case. An example of a symmetric proposal distribution for sampling the parameter space~$\Theta$ is the multivariate normal distribution, used by Szymańska et al. in~\cite{Szymanska2021} in the previously described Bayesian calibration of the non-local model of multipheroid growth. We have 
\begin{equation}\label{space_sample_parameter}
    \tilde{\theta} = \theta_{j} + Z, \;\;\; \textrm{where} \;\;\; Z \sim N(0,s \! \cdot \! \textrm{Id}),
\end{equation}
where $s$ is a step size whose proper choice is quite important in practice, as it affects the convergence rate of the algorithm~\cite{gelman1997}. Observe that the acceptance rule for a candidate state is given directly by the algorithm used. To conclude, we highlight the relationship between Eq.~\eqref{BI} and Eq.~\eqref{m_obs_log}. The parameter estimation procedure relies on the dataset $D = \{\log(r_{o}^i)\}_{i=1}^{l}$, where $l$ denotes the number of available measurements. The corresponding model predictions are given by
$\mathcal{M}(\theta) = \{\log (r(t_i))\}_{i=1}^{l}$, while the measurement uncertainty is represented by $\eta = \tilde{Z}_i$.

To numerically approximate the posterior distribution given by Eq.~\eqref{posterior}, we must compute its normalising constant, namely the integral in the denominator, to ensure that it defines a valid probability distribution. In general, a straightforward approach for evaluating this integral is the Monte Carlo method:  
\begin{equation}
    \int_0^1 f(x) dx \sim \frac{1}{n} \sum_{i=1}^{n}f(X_i),
\end{equation} 
where $X_i$ are independent random variables uniformly distributed over $[0,1]$. However, the Monte Carlo method is effective only for certain types of integrands. For integrands, where most randomly sampled points contribute negligibly and meaningful values, are concentrated in a small region, the method becomes highly inefficient. This inefficiency is particularly pronounced for functions of the form $e^{-V(\theta)},$ where $V(\theta)$ is given by Eq.~\eqref{like2}.  

Another major drawback of simple Monte Carlo sampling is its slow convergence rate, which worsens as the dimensionality of the space increases. To address this limitation, more advanced algorithms have been designed to selectively sample points in regions where the integrand is largest. By focusing computational effort on these relevant areas, such methods significantly enhance efficiency and accelerate convergence. For example, adaptive Metropolis algorithm adjusts the covariance matrix of the proposal distribution during the sampling process~\cite{Haario2001}. This allows to significantly improve the acceptance rate of the algorithm.


\subsection{Gradient flow approach}\label{subsec:gradient_flow}


Recent advances in sampling methods have introduced a shift from purely stochastic schemes, such as MCMC, to deterministic approaches based on gradient flows which  played an important role in the analysis of nonlinear PDEs and optimal transport problems~\cite{Ambrosio_Gigli_Savare_2008,Villani_2009,Santambrogio_2015}, and increasingly, in the development of deterministic sampling algorithms~\cite{CarrilloSkrzeczkowski2025,korba2020non}. These methods, such as Stein Variational Gradient Descent (SVGD) and other particle-based samplers, leverage the geometry of the underlying distribution to guide samples along trajectories that minimise a chosen divergence measure, such as the defined below Kullback–Leibler divergence.

While MCMC techniques rely on random exploration and often suffer from issues such as slow mixing and poor scalability in high dimensions, deterministic samplers aim to mitigate these limitations by exploiting the structure of the underlying distribution. In particular, they use gradient-based transport of particles toward the target distribution. This can result in algorithms that are more efficient and easier to analyse. 

A key development in this area is the formulation of sampling as a gradient flow in the space of probability measures, equipped with the Wasserstein distance. In this framework, the evolution of a probability density is viewed as the steepest descent of an energy functional with respect to this geometry. A seminal result in this direction~\cite{JKO} showed that the Fokker–Planck equation 
\begin{equation}\label{eq:Fokker-Planck}
    \partial_t\varrho={\rm div} \big(\nabla\varrho+\varrho\nabla V\big),
\end{equation}
with an initial condition $\varrho_0$ satisfying $\int_{\R^n}\varrho_0 d\theta=1$ and a sufficiently regular potential $V(\theta)$, arises as the Wasserstein gradient flow of of a free energy functional.

This interpretation has deep theoretical consequences. The Kullback–Leibler divergence between a current distribution $\varrho_{t}$ and the stationary target distribution $\varrho_{\infty}$ acts as a Lyapunov functional for the system. Given probability measures $Q$ and $P$, the Kullback–Leibler divergence is defined as
\begin{equation}
  {\displaystyle D_{\text{KL}}\big(P\parallel Q\big)=\int _{\theta\in {\Theta}} \log \left({\frac{P(d\theta)}{Q(d\theta)}}\right)\ P(d\theta),}
 \end{equation}
where ${\displaystyle{\frac{P(d\theta)}{Q(d\theta)}}}$ is the Radon-Nikodym derivative of $P$ with respect to $Q$. It is a non-symmetric measure of the difference between two probability measures, or in other words, it measures relative entropy. Serving as a basic concept for quantifying statistical distance, the Kullback–Leibler divergence has been widely used in the data mining literature.

In our setting, the potential $V(\theta)$ corresponds to the negative log-likelihood introduced in Eq.~\eqref{like2}, and the target distribution is given by 
\begin{equation*}
   \varrho_{\infty}(\theta)=\frac{e^{-V(\theta)}}{{\int_{\R^n} e^{-V(\theta)} d\theta}}. 
\end{equation*}

The Fokker-Planck equation preserves mass, has a unique stationary solution that coincides with the defined above $\varrho_{\infty}$, and Kullback–Leibler divergence (a relative entropy) between the current distribution  $\varrho_{t}$ and the stationary solution $\varrho_{\infty}$ decreases over time, ensuring convergence to equilibrium. We note that in the case of two measures $\varrho{\cal L}^n$ and $\varrho_{\infty}{\cal L}^n$ s.t. $\varrho{\cal L}^n$ is uniformly continuous with respect to $\varrho_{\infty}{\cal L}^n$, where ${\cal L}^n$ is the n-dimensional Lebesgue measure, one can write the distance in terms of densities $\varrho,\varrho_{\infty}$ 
\begin{equation}
    D_{\text{KL}}(\varrho{\cal L}^n \parallel \varrho_{\infty}{\cal L}^n):=\int_{\R^n}\psi\left(\frac{\varrho}{\varrho_{\infty}}\right)\varrho_{\infty} d\theta
\end{equation}
with $\psi(s)=s \log (s)$. 

Consequently, the Fokker-Planck equation provides a fundamental example of how the time evolution of a probability density can be interpreted as a gradient flow in the space of probability measures. The long-time asymptotic of the Fokker-Planck equation follows from the pioneering paper~\cite{Arnold2001} on the application of this relative entropy. It is well known that such $\psi$ is strictly convex and thus with help of Jensen's inequality, one can show that $\varrho_\infty$ is the unique minimiser of $D_{\text{KL}}(\varrho{\cal L}^n \parallel \varrho_{\infty}{\cal L}^n)$ among all probability densities $\varrho$. For a similar idea of the application of relative entropy for porous media see~\cite{Carrillo2000}.

More generally, if $\psi$ is  a strictly convex function, one can define a generalized relative entropy
\begin{equation}\nonumber
\mathcal{E}_\psi(\varrho \| \varrho_\infty) = \int_{\mathbb{R}^n} \psi\left( \frac{\varrho}{\varrho_\infty} \right) \varrho_\infty \, d\theta,
\end{equation}
whose time derivative satisfies
\[
\frac{d}{dt} \mathcal{E}_\psi(\varrho(t) \| \varrho_\infty) = \mathcal{I}_\psi(\varrho(t) \| \varrho_\infty),
\]
where $\mathcal{I}_\psi$ is a non-positive functional that vanishes only at equilibrium,
\begin{eqnarray*}
\mathcal{I}_\psi(\varrho(t) \| \varrho_\infty) :&=& -\int_{\R^n} |\nabla w|^2 \varrho_\infty \, d\theta, \\
w&=&F_\psi\left(\frac{\varrho(t)}{\varrho_\infty}\right), \\
F_\psi(\mu) &=& \int_1^\mu \sqrt{\psi''(s)}\, ds, \quad \mu > 0,
\end{eqnarray*}
The strict convexity of $\psi$ implies a strict monotonicity of $F_\psi$ and hence, for any probability densities $\varrho(t)$ and $\varrho_\infty$, the functional $\mathcal{I}_\psi(\varrho(t) \| \varrho_\infty)$ is non-positive and equals zero if and only if $\varrho(t) = \varrho_\infty$. This expresses the fundamental property that any relative entropy $\mathcal{E}_\psi(\varrho \| \varrho_\infty)$ decreases in time, provided that $\psi$ is strictly convex. Under suitable assumptions, one can also derive convergence rates using logarithmic Sobolev inequalities~\cite{Arnold2001}.
 
In practice, numerical implementation of gradient flow-based samplers often relies on simulating the Langevin equation,
\begin{equation}\label{eq:stochastic_ODE_FK}
{\diff} X(t) = -\nabla V(X(t))\diff t + \sqrt{2}\,  \diff B(t)\,,
\end{equation}
where $B$ is the Brownian motion, see the seminal paper~\cite{Roberts1996}, as well as more recent studies~\cite{chewi22a,dalalyan17a,Dalalyan17,Durmus2019,DKR,DM17,DM19,erdogdu21a,erdogdu22a,mousavi-hosseini23a,VW}.
Convergence analysis of Langevin-based sampling methods, rooted in the interpretation of the Fokker–Planck equation as a gradient flow in the $W_2$-Wasserstein distance, has been recently presented~\cite{Durmus2019}. A notable limitation of this approach is that, due to stochastic fluctuations, a large number of trajectories is required to obtain accurate approximations.

 To overcome this, recent efforts have focused on deterministic particle-based approximations. A major challenge in this context is that the variational structure of the Fokker–Planck equation does not extend to discrete (atomic) measures. Specifically, the density $\varrho(\mathbf{r})$ satisfies a variational principle with the corresponding free energy
\begin{equation}\nonumber
    \mathcal{F}[\varrho] = \int V(\theta)\,\varrho(d\theta) + \int \varrho(\theta) \log \varrho(\theta) \, d\theta.
\end{equation}
and the entropy term $\int \varrho(\theta) \log \varrho(\theta) \, d\theta$ becomes infinite for empirical measures composed of Dirac deltas.

One popular approach to bypass this issue is Stein Variational Gradient Descent (SVGD). Instead of directly simulating the Fokker–Planck equation, SVGD evolves the empirical distribution 
\begin{equation}\nonumber
\varrho^N(t) = \frac{1}{N} \sum_{i=1}^N \delta_{\theta_i(t)},
\end{equation}
according to a nonlocal PDE
\begin{equation} \label{eq:svgd_pde}
\partial_t \varrho = \nabla \cdot \big( \varrho K * (\nabla \varrho +  \varrho \nabla V) \big),
\end{equation}
where $K$ is a positive-definite kernel and $*$ denotes the convolution operator. 

The particle dynamics follow the system of ordinary differential equations
\begin{equation} \label{eq:svgd_ode}
\frac{d}{dt} \theta_i(t) = -\frac{1}{N} \sum_{j=1}^N \nabla K(\theta_i(t) - \theta_j(t)) - \frac{1}{N} \sum_{j=1}^N K(\theta_i(t) - \theta_j(t)) \nabla V(\theta_j(t)),
\end{equation}
with initial conditions $\theta_i(0)$ given by $\varrho^N(0) = \frac{1}{N} \sum_{i=1}^N \delta_{\theta_i(0)}$ that approximates the initial density $\varrho(0)$.

These deterministic dynamics allow for efficient sampling while preserving key properties of the underlying distribution. The convergence and stability of the particle system Eq.~\eqref{eq:svgd_ode} have been recently analysed within the framework of the $W_2$ Wasserstein distance~\cite{CarrilloSkrzeczkowski2025}. The analysis was also conducted with respect to the dual bounded Lipschitz distance (also known as the flat metric), a metric used for studying nonlocal growth and transport problems in the space of nonnegative Radon measures~\cite{Dull_Gwiazda_Marciniak-Czochra_Skrzeczkowski_2021}. The method has also been analysed in the continuous setting, where kernels were constructed that ensure exponential convergence of the solution toward the target distribution~\cite{carrillo_skrzeczkowski_warnett2024}. Another sampling approach that connects Langevin diffusion with deterministic dynamics is the so-called blob method~\cite{MR3913840}. In this context, a nonlocal equation is introduced to approximate the Fokker–Planck equation, Eq.~\eqref{eq:Fokker-Planck}. The theoretical understanding of the method remains incomplete, despite independent studies that have explored its properties~\cite{MR4858611,carrillo2024nonlocal}. However, recent works have provided the first convergence rates for the blob method in a simplified case of the porous medium equation~\cite{carrillo2025rate,amassad2025deterministic}.


In summary, gradient flow-based methods offer a powerful and theoretically well-founded alternative to stochastic sampling. Their connection to optimal transport, entropy minimisation, and PDEs provides a rich framework for both analysis and implementation. As computational methods improve, such deterministic algorithms are becoming increasingly competitive for high-dimensional sampling tasks.


\section{Discussion and Outlook}\label{sec:discussion}


Calibrating complex biological models remains a central challenge in mathematical biology, where data are often noisy, incomplete, and heterogeneous, and where model dynamics may be nonlinear, multiscale, or even chaotic. Nevertheless, successful parameter inference has enabled significant progress across various biological domains. Traditional Bayesian approaches to inverse problems provide a powerful statistical framework for parameter estimation by treating model parameters as random variables and seeking to characterise their posterior distribution. Derived using Bayes' theorem, the posterior combines prior knowledge with the likelihood of observed data, thereby allowing uncertainty to be explicitly modelled. This probabilistic treatment offers valuable insights into the inherent correlations between parameters and the reliability of the resulting estimates. Among Bayesian techniques, MCMC methods, particularly the Metropolis and Metropolis–Hastings algorithms, are widely used to approximate posterior distributions. These methods construct a Markov chain whose stationary distribution matches the target posterior by generating a sequence of samples. Starting from an initial state, typically informed by prior knowledge, the algorithm iteratively proposes new candidate states using a probability density function, often modelled as a multivariate normal distribution. The Metropolis algorithm assumes symmetry in the proposal distribution, whereas the Metropolis–Hastings algorithm allows for asymmetry, making it more general and adaptable.

Despite their theoretical appeal, MCMC based techniques can be computationally demanding, especially for high-dimensional systems or models that exhibit sensitive dependence on initial conditions. In such settings, convergence can be slow, and performance becomes highly sensitive to the choice of prior and proposal distribution. As a result, standard Monte Carlo integration may become inefficient or even intractable. These challenges have led to the development of more advanced sampling strategies aimed at improving convergence rates and reducing computational cost.

In some situations, experimental data contain such a limited amount of information that it becomes impossible to derive an analytical expression for the discrepancy between the model output and the experimental data, thereby preventing the use of standard Bayesian methods. These challenges can be addressed with the CIL method, which is a robust statistical inference framework, that compares full spatio-temporal solutions by encoding pattern similarity in the distribution of scalar features (pairwise distances). This enables its effective use even when data are incomplete, scaled, or non-stationary, making it particularly well-suited for biological models that generate complex heterogeneous structures. The accuracy of parameter identification can be significantly enhanced by incorporating multiple norms~\cite{Kazarnikov2023}. Model-specific features can also be integrated into the approach. For example, in the case of cellular automata models, various meaningful features were used alongside distance metrics to improve estimation accuracy~\cite{Kazarnikov2023a}.
 
The applicability of the CIL method extends well beyond classical Turing patterns. Its effectiveness has been demonstrated in diverse contexts, including mechano-chemical models that couple chemical processes with tissue mechanics, discontinuous {\it far-from-equilibrium} patterns produced by reaction-diffusion-ODE systems, and stochastic cellular automata. Moreover, the methodology is not restricted to stationary patterns. It remains valid for spatio-temporal structures such as chemical waves and Cahn–Hilliard phase separation dynamics. From a methodological standpoint, time-dependent dynamics of pattern formation models can be treated analogously to chaotic systems. This versatility makes the CIL method broadly applicable across a wide spectrum of spatio-temporal models.

The CIL framework has so far been validated primarily on simulated datasets and, more recently, on an experimental chemical case study (the CIMA reaction), where the underlying reaction network is well characterised and the reduced model used for inference is effectively a known approximation of the true chemistry \cite{Kazarnikov2025}. Applying CIL to real biological systems constitutes the next major challenge. Biological datasets are not only typically sparse and noisy, but also the underlying processes are only partially understood. In such contexts, mechanistic models are often hypothetical, and the primary aim is not to recover parameters of a fully established model but rather to discriminate between competing modelling hypotheses in order to identify plausible biological mechanisms. A promising direction for future CIL applications is parameter estimation in individual-based models, where model outputs can be directly confronted with image-derived spatial statistics. This is particularly relevant for biological systems in which cell-level behaviours, such as division, differentiation, and migration, drive emergent tissue-scale patterns. A concrete example is provided by lattice-based models of stem cell populations, which can be compared with high-resolution imaging datasets, such as quantitative live imaging of stem cell dynamics in zebrafish neural tissues \cite{Dray2021}. Similarly, vertex- and lattice-based models of epithelial morphogenesis, generating spatial mosaics of distinct cell fates, can be confronted with high-resolution microscopy data quantifying the spatial arrangement of labelled cell populations in developing tissues \cite{Fletcher2014,Alt2017}. A further example is provided by pattern-formation models whose simulated output can be matched to {\it in situ} hybridisation images showing spatial distributions of mRNA corresponding to one of the model variables \cite{Mercker2025}.

Complementing the likelihood formulation, we discussed deterministic alternatives to classical MCMC, grounded in the theory of gradient flows on the space of probability measures. By interpreting sampling as a transport problem minimising the Kullback–Leibler divergence, one can design algorithms, such as Fokker–Planck solvers or particle-based variational methods, that converge deterministically toward the target posterior. These approaches significantly reduce computational cost and variance, offering practical advantages in high-dimensional or computationally stiff inference tasks.

Together, these methodological advances constitute a scalable and versatile toolbox for parameter estimation, as demonstrated on a range of models describing biological processes. On example of a mathematical model of cancer colony dynamics we illustrated the utility of Bayesian inference in linking theoretical predictions with experimental observations, while rigorously accounting for noise and measurement uncertainty. The applicability of the CIL method is showcased through models that produce heterogeneous outputs, such as Turing pattern formation systems, stochastic cellular automata, and Cahn–Hilliard-type models. A key strength of the CIL approach lies in its ability to extract meaningful features even from noisy, incomplete, or scaled data. Furthermore, modern numerical techniques such as adaptive meshing, GPU acceleration, and parallel computing enable the development of computationally efficient implementations of the method, making it particularly suitable for contemporary computational biology.
Looking forward, the integration of CIL with machine-learning-based surrogates or reduced-order models has the potential to further enhance its applicability to large-scale biological systems. In turn, the gradient flow approach could provide valuable support in designing structure-preserving inference algorithms tailored to specific model classes, such as those obeying conservation laws or exhibiting symmetry-breaking bifurcations.


\section*{Acknowledgments}


The authors acknowledge support from the Excellence Initiative – Research University program (2020–2026) at the University of Warsaw, Poland. P.G. and Z.S. were partially supported by the National Science Centre, Poland, grant UMO-2024/54/A/ST1/00159. A.M.C. and A.K. are supported by the European Research Council (ERC) under the European Union’s Horizon 2020 research and innovation programme (synergy project PEPS, no. 101071786), and by the Deutsche Forschungsgemeinschaft (DFG) within the Collaborative Research Centre SFB1324 (B05) and within Germany’s Excellence Strategy EXC 2181/1 - 390900948 (the Heidelberg STRUCTURES Excellence Cluster). 

\bibliography{sn-bibliography}  

\end{document}